\documentclass[pdflatex,sn-nature]{sn-jnl}

\usepackage{graphicx}%
\usepackage{multirow}%
\usepackage{amsmath,amssymb,amsfonts}%
\usepackage{amsthm}%
\usepackage{mathrsfs}%
\usepackage[title]{appendix}%
\usepackage{xcolor}%
\usepackage{textcomp}%
\usepackage{manyfoot}%
\usepackage{booktabs}%
\usepackage{algorithm}%
\usepackage{algorithmicx}%
\usepackage{algpseudocode}%
\usepackage{listings}%

\usepackage{pdflscape}
\usepackage{adjustbox}

\usepackage{tabularx}
\usepackage{longtable}
\usepackage{enumitem}
\usepackage{ragged2e}

\usepackage{xcolor,colortbl}
\definecolor{mygray}{gray}{0.88}
\usepackage{hhline}

\usepackage{subcaption}

\theoremstyle{thmstyleone}%
\theoremstyle{thmstyletwo}%

\theoremstyle{thmstylethree}%

\begin{document}

\title[NM-Brief-Communication]{Fine-grained Alignment of Large Language Models for General Medication Recommendation without Overprescription}


\author[1]{\fnm{Zihao Zhao} 
}\email{zihaozhao1998@gmail.com}
\equalcont{These authors contributed equally to this work.}

\author[1]{\fnm{Chenxiao Fan} 
}\email{simonfan@mail.ustc.edu.cn}
\equalcont{These authors contributed equally to this work.}

\author[2]{\fnm{Junlong Liu} 
}\email{pingwu.ljl@alibaba-inc.com}

\author[2]{\fnm{Zheng Wang} 
}\email{wz388779@alibaba-inc.com}

\author[3]{\fnm{Xiangnan He} 
}\email{xiangnanhe@gmail.com}

\author*[1]{\fnm{Chongming Gao} 
}\email{chongminggao@ustc.edu.cn}

\author[4]{\fnm{Juan Li} 
}\email{huamuzi1999@hotmail.com}

\author*[3]{\fnm{Fuli Feng} 
}\email{fulifeng93@gmail.com}

\affil[1]{\orgdiv{School of Information Science and Technology}, \orgname{University of Science and Technology of China}, \orgaddress{\street{100 Fuxing Road}, \city{Hefei}, \postcode{230088}, \state{Anhui}, \country{China}}}

\affil[2]{\orgname{Alibaba Group}, \city{Beijing},  \country{China}}

\affil[3]{\orgdiv{School of Artificial Intelligence and Data Science}, \orgname{University of Science and Technology of China}, \orgaddress{\street{100 Fuxing Road}, \city{Hefei}, \postcode{230088}, \state{Anhui}, \country{China}}}

\affil[4]{\orgdiv{Department of Anesthesiology, The First Affiliated Hospital of USTC, Division of Life Sciences and Medicine}, \orgname{University of Science and Technology of China}, \orgaddress{\street{100 Fuxing Road}, \city{Hefei}, \postcode{230088}, \state{Anhui}, \country{China}}}




\abstract{Large language models (LLMs) holds significant promise in achieving general medication recommendation systems owing to their comprehensive interpretation of clinical notes and flexibility to medication encoding. We evaluated both general-purpose and medical-specific LLMs for medication recommendations, showing their unsatisfactory precision and severe overprescription. To address this, we introduce Language-Assisted Medication Recommendation, which tailors LLMs for medication recommendation in a medication-aware manner, improving the usage of clinical notes. Fine-tuning LLMs with this framework can outperform existing methods by more than 10\% in internal validation and generalize across temporal and external validations. Furthermore, the model maintains high accuracy when encountering out-of-distribution medication.}




\maketitle

\section*{Introduction}

The growing importance of medication recommendation systems~\cite{medrec_survey} in clinical decision-making is amplified by the rapid evolution of Electronic Health Records (EHRs)~\cite{cowie2017electronic}. 
Their ability to prescribe precise and efficacious drug combinations offers a compelling pathway to enhance the efficiency of pharmacotherapy, especially given the ever-expanding pharmacopoeia that often extends beyond the immediate areas of specialization for many primary care physicians. 
Furthermore, these systems hold considerable promise for advancing equitable access to quality healthcare, particularly in resource-constrained regions, such as South Asia and Africa~\cite{foreman2018forecasting}. Consequently, the development of reliable and general medication recommendation systems becomes a critical objective for both national and international health infrastructures\footnote{https://www.fda.gov/regulatory-information/search-fda-guidance-documents/regulatory-considerations-prescription-drug-use-related-software?utm\_source=chatgpt.com}.

Existing medication recommendation systems~\cite{molerec, raremed} exhibit limited generalizability due to their dependence on ID-based medication encoding~\cite{gamenet, safedrug, cognet} and the underutilization of clinical notes. ID-based encoding fundamentally constrains these approaches in adapting to expanding pharmacopoeia and heterogeneous coding practices found across institutions. Their prevailing focus on structured data sourced from EHR systems prevents them from capturing nuanced and multifaceted clinical profiles of patients~\cite{goh2021artificial} since nearly 80\% of clinical information in EHR systems resides in unstructured notes without standardized schemas~\cite{martin2014big}. Addressing these fundamental limitations requires the development of medication recommendation frameworks with flexible encoding schemes and comprehensive interpretation of clinical notes.

Large language models~\cite{LLM_medicine_nature} hold significant promise in achieving the target, given their proficiency in processing free-text information and strong generalization ability. Medical-domain LLMs have demonstrated remarkable performance in biomedical tasks~\cite{medpalm, tang2023evaluating}, such as the expert-level score of MedPaLM-2~\cite{medpalm2} in the United States Medical Licensing Examination. Nevertheless, our evaluation reveals that both general and medical LLMs underperform in medication recommendation, when tasked to recommend a medication to a given patient or not with their textual descriptions as inputs. More troublingly, these LLMs exhibit a pronounced tendency to overprescribe. GPT-4~\cite{gpt4} recommends over 80 medications per patient on average --- over three times the volume of practicing physicians. 
In absence of rigorous calibration, deploying these LLMs for medication recommendation risks driving up healthcare expenditures, elevating the incidence of adverse drug events, and hastening the emergence of antimicrobial resistance~\cite{ferrara2017antibiotic}.

In this research, we introduce Language-Assisted Medication recOmmendation (LAMO), a novel approach that leverages Low-Rank Adaption (LoRA)~\cite{lora} to tailor LLMs for medication recommendation. To overcome overprescription, LAMO adopts a mixture-of-expert strategy with different LoRAs making decisions for different medicines. To control the computation overhead, LAMO groups medicines into clusters and let medicines in a group share a LoRA. By reorganising a representative EHR dataset MIMIC-III~\cite{mimic-iii} into an instruction tuning manner, we finetune a representative open-source LLM LLaMA~\cite{llama2}, obtaining a general medication recommendation model. The LAMO model surpasses state-of-the-art (SOTA) methodologies in the internal validation on MIMIC-III, the temporal validation on MIMIC-IV~\cite{mimic-iv}, and the external validation eICU~\cite{eicu}, a multi-center dataset comprising data from over 200 hospitals. In addition to the generalization capabilities across diverse temporal and hospital contexts, LAMO exhibits remarkable accuracy on out-of-distribution medications absent from the training dataset.

\section*{Methods}


\subsection*{Structured EHR Representation for LLMs}
We curated structured inputs for large language models (LLMs) by extracting and summarizing key clinical factors from raw discharge summaries (Supplementary Table~S1). Each note was parsed into four core components: \textbf{History of Present Illness}, \textbf{Past Medical History}, \textbf{Allergies}, and \textbf{Medications on Admission}. These fields collectively capture the patient's current condition, comorbidities, medication history, and known sensitivities.

To construct each component, we used standardized prompts with GPT-3.5 (Supplementary Table~S2) to segment and extract relevant spans from raw notes, followed by summarization to satisfy input length constraints. The resulting outputs served as structured representations for downstream LLM processing  (Supplementary Table~S3, Figure 1a). For patients with prior visits, we incorporated historical diagnoses, procedures, demographics, and previously prescribed medications, while omitting full clinical notes to limit input length. Discharge medications were explicitly excluded to prevent label leakage. When notes exceeded context limits and required segmentation, we applied consolidation prompts to merge overlapping or partial extractions into a unified form (Supplementary Table~S4).

\subsection*{LLM Fine-Tuning for Medication Recommendation}
We adopted \textbf{LLaMA-2-7B} as our backbone model due to its strong performance and public availability. The model was aligned to the medication recommendation task through instruction tuning.

\textbf{Instruction Tuning.} Each training instance was formulated using an instruction-based format. We defined a \textit{Task Instruction} describing the recommendation task, paired with \textit{Task Input} fields capturing the structured clinical context and a candidate medication. The \textit{Instruction Input} concatenates these, while the \textit{Instruction Output} is a binary label indicating whether the medication was prescribed. The formulation enables LLMs to learn context-sensitive medication decisions (Figure 1a).

\textbf{Parameter-Efficient Fine-Tuning.} We employed Low-Rank Adaptation (LoRA) for fine-tuning, enabling efficient training with limited computational overhead. LoRA injects trainable low-rank matrices into frozen Transformer layers. We trained separate LoRA adapters for distinct medication groups (Figure 1b). During inference, all relevant adapters were activated to collectively generate the medication recommendation list.

In fine-tuning the LAMO model, we employed a learning rate of 5e-4 in conjunction with an inverse square root scheduler, utilizing a batch size of 64. The rank of LoRA adapters was specified as 8, with alpha set to 16, dropout at 0.05, and target modules comprising ``q\_proj'' and ``v\_proj''. F1 score evaluation on the validation set was conducted every 32 batches, with training halted upon ten consecutive assessments showing no improvement in accuracy.
For model inference, a temperature of 0 was uniformly applied to all models to ensure experiment reproducibility. All other parameters were maintained at their default settings.

\section*{Results}

\textbf{Unsatisfactory Accuracy and Overprescription.}
We evaluate eight LLMs on the MIMIC-III~\cite{mimic-iii} dataset to assess their performance in medication recommendation (detailed prompts in Supplementary Tables S5–S10). GPT-4 demonstrates the highest accuracy among these models, achieving an F1 score of 0.3542, which significantly lags behind that of traditional SOTA methods (Figure 2a). Moreover, these models tend to recommend a notably higher number of medications for patients, with LLaMA-2 and LLaMA-3~\cite{llama2} being particularly inclined to suggest nearly all candidate medications for each patient (Figure 2b). ChatGLM3 performs especially poorly, frequently failing to generate any medication recommendations across patients, resulting in extremely low accuracy. These results underscore the significant gap between general medical reasoning and medication recommendation, emphasizing the necessity for further investigation into the specific challenges and limitations faced by LLMs in medication recommendation.

\textbf{Insufficient Medication Knowledge.}
To uncover the reason for the unsatisfactory performance, we conduct a knowledge test on a comprehensive biomedical knowledge graph PrimeKG~\cite{primekg}, evaluating whether these LLMs can recognize the relation between diseases and medicines (detailed prompts in Supplementary Table S11–S14). Most tested LLMs underperform with accuracy below 80\% (Figure 2c, Supplementary Table S15-S22),  revealing significant limitations in both general and medical LLMs about nuanced medical relationships. This result highlights the importance of further fine-tuning LLMs for medication recommendation to bridge this knowledge gap.

\textbf{Effectiveness of LAMO Fine-tuning.}
We conducted internal validation, temporal validation, and external validation for LAMO against general LLMs, medical LLMs, conventional medication recommendation models (Supplementary Table S23). 
\textbf{a) Internal Validation.} LAMO outperforms all baseline methods in terms of recommendation accuracy on the test set of MIMIC-III with a moderate number of recommendations for patients, akin to domain-specific traditional models such as SafeDrug~\cite{safedrug} and MoleRec~\cite{molerec} (Figure 2d, Supplementary Table S24). This result suggests that LAMO activates the capabilities of LLMs in analyzing the clinical conditions in the unstructured notes, addressing the issue of overprescription.
\textbf{b) Temporal Validation.} We then compared LAMO with representative baselines on the MIMIC-IV~\cite{mimic-iv} dataset, which follows ICD-10 to encode diseases and procedures updated from ICD-9 used in MIMIC-III. Despite the general performance decrease over time, LAMO showcases superior performance across all groups with an increasing improvement ratio compared to the strongest baseline RAREMed (Figure 2e), demonstrating its flexibility to the evolution of medical coding standards and strong temporal generalization ability.
\textbf{c) External Validation.} We further conducted evaluation on the eICU Collaborative Research Database, a multi-center dataset collected from a diverse set of hospitals across the United States, encompassing different patient populations and clinical practices. 
LAMO model significantly outperforms the other models (Figure 2f), indicating its robustness in diverse clinical environments and the necessity of fine-tuning for broader application of medication recommendation systems. 

\textbf{Generalization to New Medications.}
We conducted a zero-shot transfer testing to evaluate the generalization ability of LAMO, where the LLM is fine-tuned on some source medications and tested on different target medications.
LAMO demonstrated close recommendation accuracy between source and target medications within the same medication group (Supplementary Fig. S1), indicating a strong transferability across similar medications. 
This result shows the effectiveness of LAMO in learning general patterns across similar drugs, allowing it to adapt to new medications not present in the training set.

\textbf{In-depth Analysis of LAMO.} We further investigated the rationality of components designed in LAMO on the MIMIC-III dataset, including the concise title, input factors, and group-wise LoRA.
\textbf{a) Using Concise Titles.}
To validate the rationality of processing clinical codes from the original titles into concise ones, we compared LAMO with several variants using different title formats. LAMO with concise titles outputs all variants significantly (Figure 3a), validating the superiority of the concise titles in capturing essential semantic information while maintaining conciseness and relevance for comprehension by LLMs.
\textbf{b) Contribution of Input Factors.}
To evaluate the contribution of different input factors of LAMO from the EHR system, we evaluated several variants of LAMO by removing history, note, code, disease, or procedure, respectively. All variants encounter a significant performance drop as compared to the full LAMO (Figure 3b, 3c), validating the importance of a comprehensive understanding of patient information, especially the unstructured clinical note. A case study (Figure 3d) on psychotropic medications (such as antidepressants) further validates it.
\textbf{c) Group-wise LoRA.}
To investigate the impact of sharing LoRA across medications in a group, we evaluated LAMO with different group numbers. Although decreasing the number of LoRAs causes a performance drop, LAMO with 5 or 10 LoRAs remains competitive and outperforms all traditional methods (Figure 3e). This suggests the potential for developing LLMs tailored for this task with improved efficiency.

\section*{Discussion}

We systematically evaluated eight general and medical LLMs on a medication recommendation benchmark derived from EHRs, revealing substantial limitations in accuracy and a pronounced tendency toward overprescribing. Further analysis indicated that current LLMs lack adequate understanding of drug–disease associations and nuanced pharmacological relationships. To address these gaps, we introduced LAMO, a tuning-based framework that clusters medications and trains dedicated LoRA adapters for each group. LAMO consistently outperformed all baselines in F1 score and exhibited strong generalization across temporal shifts and datasets, underscoring its clinical applicability.

Our findings hold implications at both technical and societal levels. By reducing overprescription and improving recommendation accuracy, LAMO could help minimize adverse drug events, which contribute to approximately 3.5\% of hospital admissions~\cite{bouvy2015epidemiology}. In low-resource regions, it may serve as a decision-support tool for non-specialists, enabling timely and accurate treatment, especially for diseases with high burdens such as malaria and tuberculosis. Moreover, the framework could be extended to neglected diseases like Chagas or leishmaniasis, offering a scalable solution to bridge expertise gaps and promote healthcare equity. More broadly, our work contributes to the advancement of AI-driven medical reasoning and supports progress toward global health goals.

While effective, our method currently relies on supervised fine-tuning (SFT), which may not scale efficiently for large drug vocabularies. The approach also encourages rote memorization of drug-knowledge pairs, limiting the model's ability to reason about synergistic effects or safety constraints such as adverse drug-drug interactions (DDIs). Future work could incorporate pre-learned pharmacological representations to inject structured drug knowledge more efficiently, and adopt list-wise alignment strategies to help LLMs learn to recommend coherent sets of medications, rather than treating each drug independently.

\backmatter

\section*{Acknowledgements}
This research was supported by the advanced computing resources provided by the Supercomputing Center of the USTC.

\section*{Author Contributions}
Z.Z. and C.F. contributed to the methodology, data analysis, and writing of the manuscript. J. L. and Z.W. contributed to the provision of computational resources, implementation support, and validation of experimental results. X.H. contributed to the revision of the manuscript. C.G. and F.F. supervised the project and contributed to the conceptualization, methodology, and revision of the manuscript.

\section*{Competing interests}
The authors declare no competing interests.

\section*{Figure Legends}

\noindent \textbf{Figure 1.}
Overview of the data processing and the fine-tuning process of LAMO.

\medskip
\noindent \textbf{Figure 2.}  
Evaluation of LLMs for medication recommendation.  
(a) F1 scores of mainstream LLMs for medication recommendation, reflecting overall predictive accuracy.  
(b) Average number of medications recommended per patient by each LLM.  
(c) Accuracy of LLMs in knowledge-based medication recommendation under Setting 1, as detailed in Supplementary Table~S15–S22.  
(d) Overall performance comparison between traditional methods and LLM-based approaches, including both F1 scores and the number of recommended medications.  
(e) Temporal validation results on the MIMIC-IV dataset, where patients are divided into four cohorts based on admission time. LAMO is evaluated against representative baselines with emphasis on its performance gain over the strongest baseline (RAREMed) in each cohort.  
(f) External validation results on the eICU dataset.

\medskip
\noindent \textbf{Figure 3.}  
Ablation and factor analysis of LAMO's performance.  
(a) Recommendation precision of LAMO under different representation strategies of diseases and procedures.  
(b) Recommendation precision LAMO variants with the removal of different input factors.  
(c) Comparative accuracy for psychiatric versus general medications. ``w/o His. \& Note'' denotes the removal of all historical and current note information. MoA refers to Medications on Admission, HPI to History of Present Illness, and PMH to Past Medical History.  
(d) Medication-specific performance degradation is observed when corresponding clinical factors are excluded from the input, particularly for psychiatric drugs.
(e) Model performance with varying numbers of LoRA adapters.




\section*{Data availability}


\begin{itemize}
    \item The MIMIC-III dataset is available at \url{https://physionet.org/content/mimiciii/1.4/}. 
    \item The MIMIC-IV dataset, used for temporal validation, is available at \url{https://physionet.org/content/mimiciv/2.0/}. 
    \item The eICU Collaborative Research Database, utilized for external validation, is available at \url{https://physionet.org/content/eicu-crd/2.0/}. 
    \item The PrimeKG dataset, used for knowledge evaluation, is available at \url{https://dataverse.harvard.edu/dataset.xhtml?persistentId=doi:10.7910/DVN/IXA7BM}.
\end{itemize}
Further details and comprehensive statistics of these datasets are provided in Supplementary Table~S25.

\section*{Code availability}
The code to reproduce main and supplementary analyses is available at \url{https://github.com/zzhUSTC2016/LAMO}.


\begin{landscape}
\newpage
\thispagestyle{empty}
\begin{figure}[!t]
\centering
\vspace{-20mm}
\hspace{-40mm}
\begin{adjustbox}{width=1.8\textwidth, center}
    \includegraphics[width=1\textwidth]{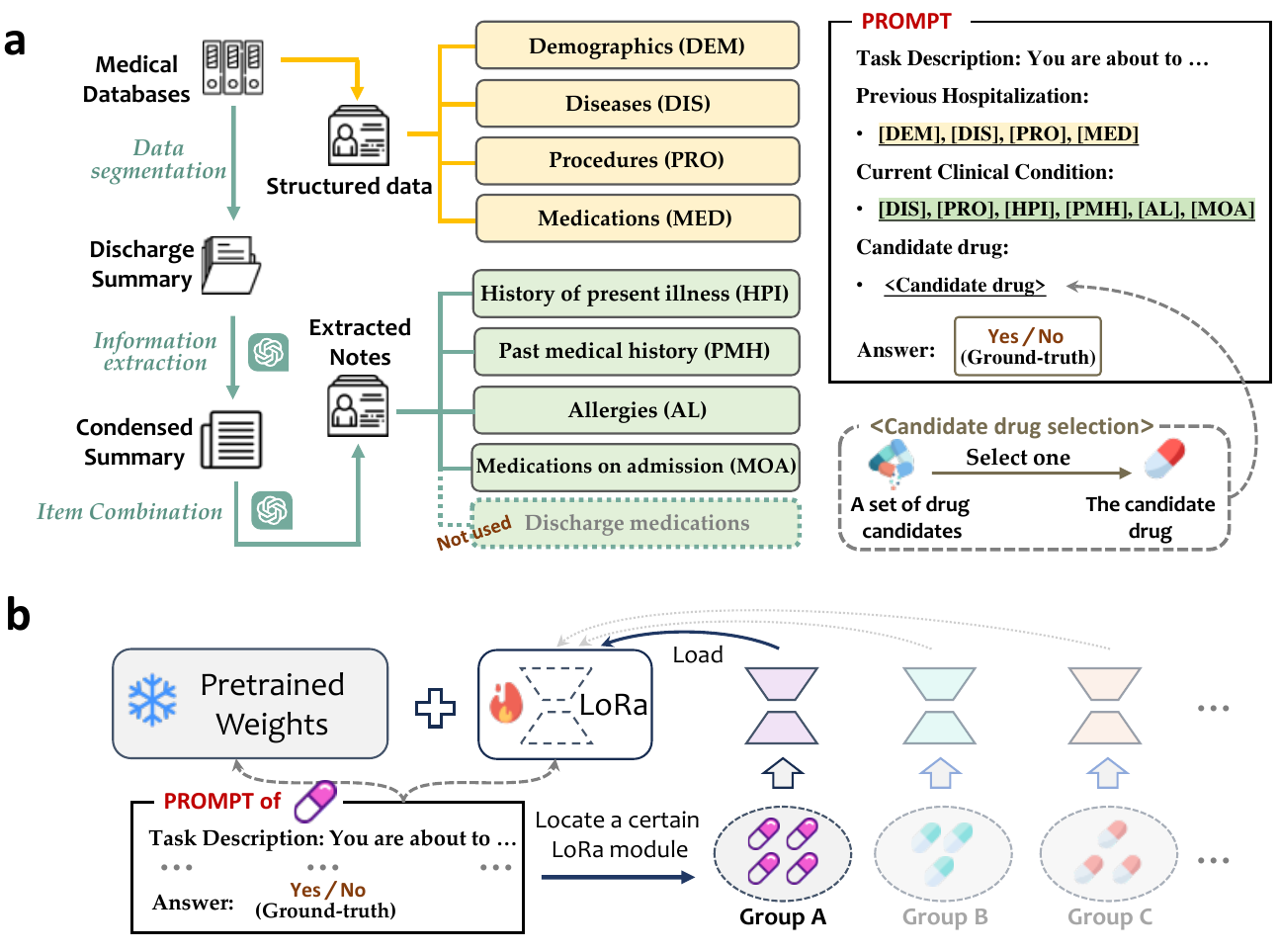}
\label{fig:model}
\end{adjustbox}
\end{figure}
\end{landscape}

\begin{landscape}
\newpage
\thispagestyle{empty}
\begin{figure}[!t]
\centering
\vspace{-30mm}
\hspace{-30mm}
\begin{adjustbox}{width=1.5\textwidth, center}
    \includegraphics[width=\textwidth]{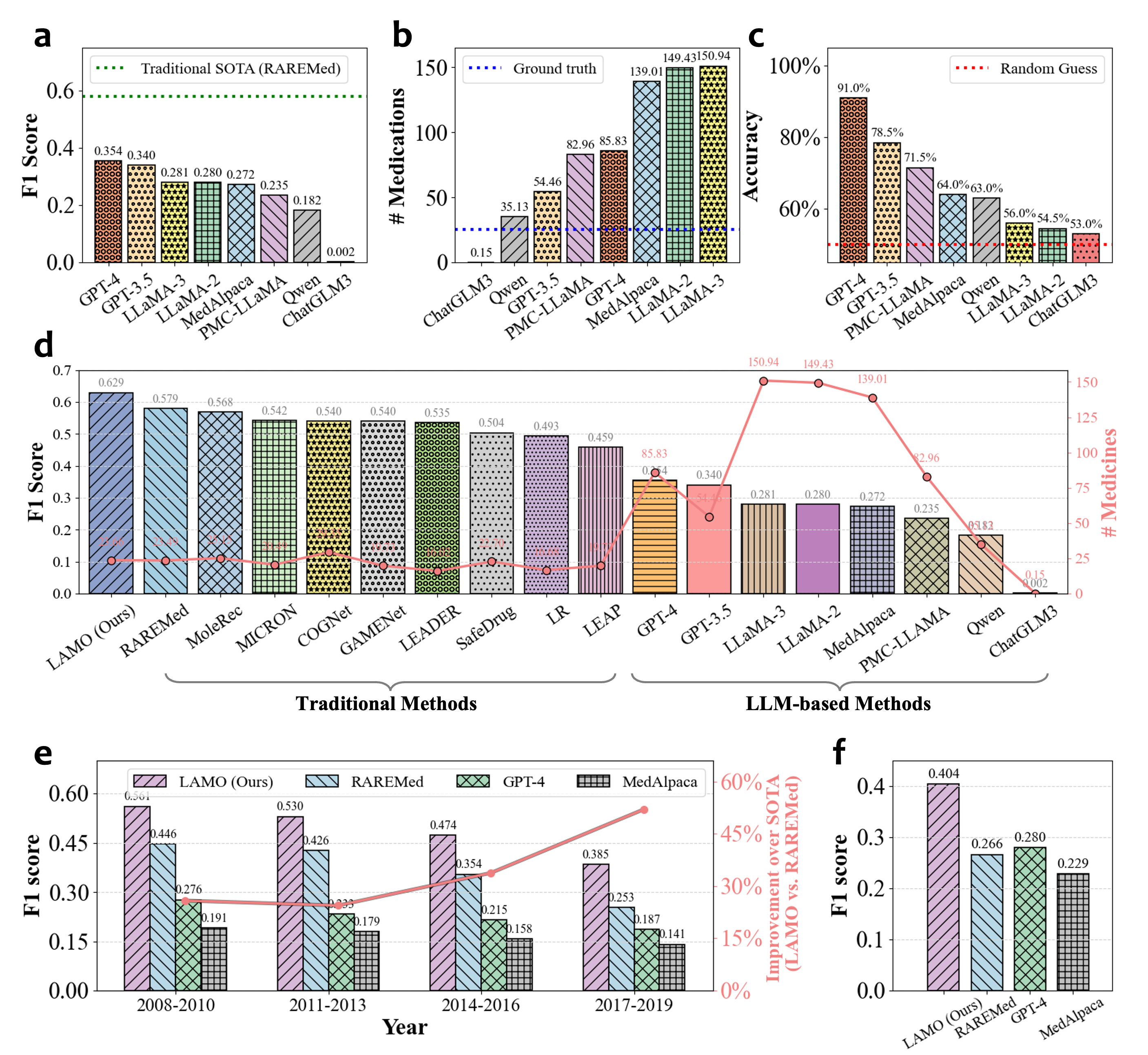}
\label{fig1}
\end{adjustbox}
\end{figure}
\end{landscape}

\begin{landscape}
\newpage
\thispagestyle{empty}
\begin{figure}[!t]
\centering
\vspace{-10mm}
\hspace{-40mm}
\begin{adjustbox}{width=1.9\textwidth, center}
    \includegraphics[width=\textwidth]{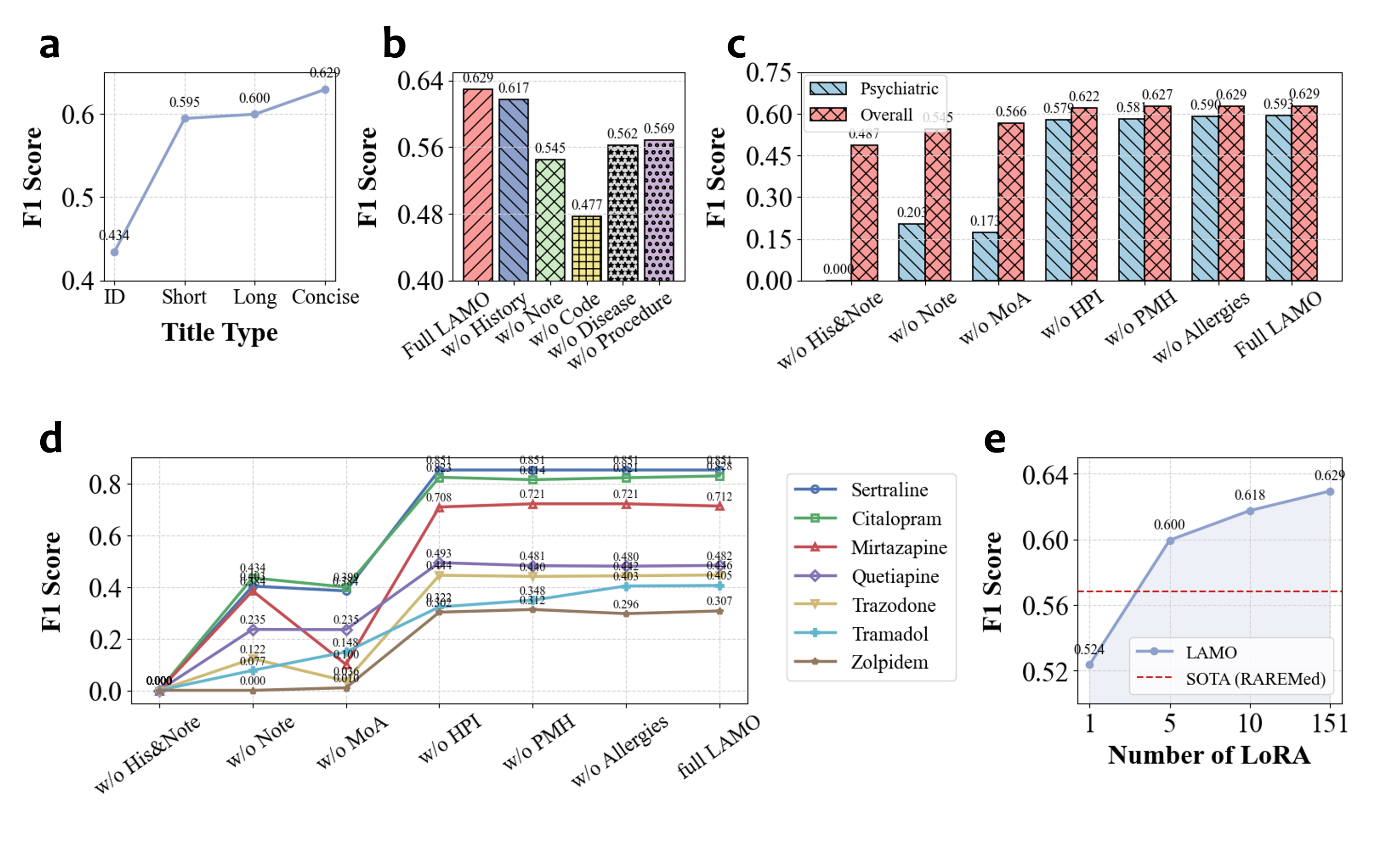}
\label{fig2}
\end{adjustbox}
\end{figure}
\end{landscape}

\clearpage

\renewcommand{\thetable}{S\arabic{table}}
\renewcommand{\thefigure}{S\arabic{figure}}
\setcounter{table}{0}   
\setcounter{figure}{0}  


\newpage
\begin{table}
    \centering
    \caption{An example of a raw clinical note. Sensitive information has been omitted to ensure confidentiality. Note that the subheadings in bold are not consistent across all patients. For example, ``Medications on Admission'' could be labeled as ``Admission Medications'', ``Medications'', etc. We have reformatted the original text to make it clearer.}
    \label{SI_tab:raw_note}
    \begin{tabular}{p{0.95\textwidth}}
    \hline
    \textbf{Admission Date}:  [**XXXX-X-XX**]  \\
    \textbf{Discharge Date}:   [**XXXX-X-XX**] \\
    \textbf{Service}: MEDICINE \\
    \textbf{Allergies}: Patient recorded as having No Known Allergies to Drugs\\
    \textbf{Attending}: [**First Name3 (LF) 7934**]\\
    \textbf{Chief Complaint}: Altered MS [**First Name (Titles) **] [**Last Name (Titles) **]\\
    \textbf{Major Surgical or Invasive Procedure}: None\\
    \textbf{History of Present Illness}:\\
    Age over 90 yo M w/ PMH multiple mylemoa, with Plasmacytoma of leftclavicle dx in [**6-18**] s/p xrt (last ~[**8-26**]) with recent admission([**Date range (3) 110715**]) for dehydration and PNA presents with one dayof altered MS [**First Name **] [**Last Name **]. Pt recently returned home fromrehab 5 days ago......
    
    In ED, found to be hypotensive to 90's systolic which was responsive to fluids.  Also febrile, w/ T 101.  Labs notable for Hct down to 25 (from baseline 28-30), Cr elevated to 2.4 (from baseline 1.1-1.3), elevated trop/CKs.  Pt was trace Guiac + on exam ......\\
    \textbf{Past Medical History}:\\
    The patient was referred to Dr. [**Last Name**] for a left clavicular mass in [**6-18**]. The patient had a history of fx in the left humerous. In early [**2196**], he developed a mass in his left shoulder. At first this was thought to be a deformity post-fracture but it continued to grow so it was decided to biopsy it. On needle biopsy [**2196-6-23**] the mass was found to be a plasmacytoma ......\\
    \textbf{PAST SURGICAL HISTORY}:\\
    1. s/p appendectomy\\
    2. s/p status post inguinal hernia repair x2\\
    \textbf{Social History}:
    The patient is married, lives in his own home.  His son ……\\
    \textbf{Family History}:
    Noncontributory\\
    \textbf{Physical Exam}:\\
    Vitals - T 99.1 (axillary), HR 113, BP 84/40 -\textgreater  SBP 62, RR 18, O2 94\% 2L NC General - pt moaning, non-responsive to verbal commands HEENT - CVS - distant heart sounds ......\\
    \textbf{Pertinent Results}: \\
    2196-9-27 02:30PM   PT-13.6* PTT-33.5 INR(PT)-1.2* \\
    \textbf{Brief Hospital Course}:\\
    Pt is a [**Age over 90 **] yo man with MMP including multiple myeloma, with Plasmacytoma of left clavicle , s/p recent admission for dehydration and pna ……\\
    \textbf{Medications on Admission}:\\
    Prednisone 5mg QD\\
    Protonix 40 QD\\
    Isosorbide mononitrate CR 30mg QD\\
    Tramadol PRN 
    ......\\
    \textbf{Discharge Medications}: None\\
    \textbf{Discharge Disposition}: Expired\\
    \textbf{Discharge Diagnosis}: 1. Septic Shock 2. Pneumonia\\
    \textbf{Discharge Condition}: Expired\\
    \textbf{Discharge Instructions}: None\\
    \textbf{Followup Instructions}: None\\
    \hline
    \end{tabular}
\end{table}

\begin{table}
    \centering
    \caption{Detailed prompt template for extracting key factors from discharge summary, where the contents in ``\{\{\}\}'' are replaced by specific elements when fed to models.}
    \label{SI_tab:extract}
    \begin{tabular}{p{0.95\textwidth}}
    \hline 
    \rowcolor{mygray}
    \textbf{Prompt template:} \\
        Please summarize specific sections from a patient's discharge summary: 1. HISTORY OF PRESENT ILLNESS, 2. PAST MEDICAL HISTORY, 3. ALLERGIES, 4. MEDICATIONS ON ADMISSION 5.DISCHARGE MEDICATIONS. Ignore other details while in hospital and focus only on these sections.\newline 
        \newline
        Output template:\newline
        HISTORY OF PRESENT ILLNESS:\newline
        (Language summary as short as possible)\newline \\
        PAST MEDICAL HISTORY:\newline
        (Language summary as short as possible)\newline \\
        ALLERGIES:\newline
        (A series of allergies names, separated by commas, does not require any other information)\newline \\
        MEDICATIONS ON ADMISSION:\newline
        (A series of drug names, separated by commas, remove dosage information. Maybe None.)\newline \\
        DISCHARGE MEDICATIONS:\newline
        (A series of drug names, separated by commas, remove dosage information. Maybe None.)\newline \\
        Note:  \{\{note\}\} \newline
        Summarize result in five aspects in a concise paragraph without any other words:\\
    \hline
    \rowcolor{mygray}
    \textbf{An example of output:}\\
       HISTORY OF PRESENT ILLNESS: The patient is a male over 90 years old who fell down 10 steps at home and was found unattended for 2 hours. He experienced vomiting and complained of headache and dizziness. He was transferred to the hospital for further evaluation and treatment.\\\\

       PAST MEDICAL HISTORY: The patient has a history of angina, a 7cm AAA scheduled for endovascular repair, hypertension, atrial fibrillation, peripheral vascular disease, aortic valve replacement, pacemaker placement, RLE bypass, and cataracts.\\\\

       ALLERGIES: No known allergies to drugs.\\\\

       MEDICATIONS ON ADMISSION: Coumadin, digoxin, atenolol, zantac, isosorbide, clonopin, hctz, altace, norvasc, and nitroglycerin.\\\\

       DISCHARGE MEDICATIONS: Isosorbide dinitrate, ramipril, amlodipine besylate, docusate sodium, atenolol, and ursodiol.\\
    \hline
    \end{tabular}
\end{table}

\newpage
\begin{table}
    \centering
    \caption{Illustration of the \textit{history} and the \textit{input} in the context of patient hospitalizations, where the \textit{history} encompasses the clinical condition recorded during the patient's last visit, if applicable (excluding the first visit), and the \textit{input} summaries the clinical status observed on the patient's current visit. Note that the Age and Gender terms are omitted in the \textit{input} as they are contained in the text of History of present illness. The contents in ``\{\{\}\}'' are replaced by specific elements when fed to models.}
    \label{SI_tab:input_example}
    \begin{tabular}{p{0.12\textwidth} | p{0.85\textwidth}}
    \hline
        \rowcolor{mygray}
        \textbf{Categories}& \textbf{Example of the \textit{history}:} \\
        \hhline{-|-}
        \multirow{1.5}{*}{\textbf{ Demon-}} & \textbf{Age}: 82, \\
        \multirow{0.7}{*}{\textbf{ graphics}} & \textbf{Gender}: female,\\
        \hhline{-|-}
        & \textbf{Diagnoses}: [Closed Transcervical Fracture of Neck of Femur, Malignant Hypertensive End Stage Renal Disease, Fall, Gout, Type II Diabetes], \\
        \textbf{ Clinical} & \textbf{Procedures}: [HD, Femur Reduction and Internal Fixation], \\
        \textbf{ \hspace{0.35pt} Codes} & \textbf{Medications}: [Acetaminophen, Oxycodone, Ranitidine, Warfarin, Dopamine, Cefazolin, Ondansetron, Lansoprazole, Cyanocobalamin, Biotin, Ascorbic acid, Riboflavin, Thiamine, Folic acid, Pyridoxine, Niacin, Pantothenic acid, Allopurinol].\\
         \hline
        \rowcolor{mygray}
        \textbf{Categories} & \textbf{Example of the \textit{input}:} \\
         \hhline{-|-}
        \multirow{5.7}{*}{\textbf{ Clinical}}& \textbf{Diagnoses}: [Ovarian Cancer, Pleural Effusion, Retroperitoneal and Peritoneal Secondary Malignancy, Urinary Tract Infection, Acute Kidney Injury, Dehydration, Coagulation Factor Deficiency, Umbilical Hernia, Hypertension, Essential, Breast Cancer History, Acquired Hypothyroidism], \\
        \multirow{-2.3}{*}{\textbf{ \hspace{0.35pt} Codes}}& \textbf{Procedures}: [Venous Catheterization, Blood Transfusion, Nutritional Infusion, Thoracic Tap, Cranial and Peripheral Nerve Division, Lymph Node Excision, Umbilical Hernia Repair, Oophorectomy, Peritoneal Biopsy],\\
        \hhline{-|-}
        \multirow{7.2}{*}{\textbf{ Clinical}} & \textbf{History of present illness}: The patient is a 66-year-old female with a history of breast cancer and recent onset ascites and pelvic mass. She presented with worsening abdominal discomfort, poor oral intake, and dyspnea.\\
        \multirow{3.25}{*}{\textbf{ \hspace{0.35pt} Notes}} & \textbf{Past medical history}: The patient has a history of breast cancer, hypertension, hypothyroidism, tubal ligation, and a previous metacarpal fracture.\\
        & \textbf{Allergies}: [codeine],\\
        & \textbf{Medications on admission}: [Lisinopril, Effexor, Levoxyl, Tamoxifen, Fosamax], \\
        & \textbf{Candidate drug}: \{\{name of the candidate drug\}\}.\\
    \hline
    \end{tabular}
\end{table}

\newpage

\begin{table}
    \centering
    \caption{In instances where a note exceeds the input capacity of GPT-3.5 and undergoes segmentation, the resultant segments may yield multiple distinct components. To consolidate these extracted results effectively, we have developed refined prompts aimed at seamlessly amalgamating the segmented information. Where the contents in ``\{\{\}\}'' are replaced by specific elements when fed to models.}
    \label{SI_tab:connect}
    \begin{tabular}{p{0.95\textwidth}}
    \hline
    \rowcolor{mygray}
    \textbf{History of Present Illness:}
    \\
         I'll provide you with an input containing the history of present illness for a patient. Your task is to:\newline
         1.Retain the descriptions of the patient's history of present illness before admission and on admission, while removing all descriptions after admission and at discharge.\newline
         2.Consolidate the text to produce a concise output.\newline\newline
         Input:  \{\{part 1\}\} + \{\{part 2\}\} + ... + \{\{part n\}\}
         \newline
         You only need to answer the refined results, no other explanation is needed!\newline
         Output:
         \\
    \hline

    \rowcolor{mygray}
    \textbf{Past Medical History:} \\
        I'll provide you with input containing a patient's past medical history. I need you to consolidate the text and output a concise summary.\newline \newline
        Input:  \{\{part 1\}\} + \{\{part 2\}\} + ... + \{\{part n\}\}
        \newline
        You only need to answer the refined results, no other explanation is needed!\newline
        Output: \\
    \hline

    \rowcolor{mygray}
    \textbf{Allergies:} \\
        I'm going to give you an input, which is a bunch of text and some plus signs. I need you to extract all the drug names for me from each input, and output the corresponding list.\newline\newline
        Here are some of the input and output sample:\newline
        Input1: No Known Allergies to Drugs.  +  None mentioned.\newline
        Output1: []\newline
        Input2: None mentioned.  +  The patient is allergic to cefazolin and penicillins.\newline
        Output2: [cefazolin, penicillins]\newline\newline
        Now you need to provide the corresponding output of input3, without any other words:\newline
        Input3:  \{\{part 1\}\} + \{\{part 2\}\} + ... + \{\{part n\}\}\newline
        You only need to output a list!\newline
        Output3: \\

    \hline
    \rowcolor{mygray}
    \textbf{Medication on Admission:} \\
        I'm going to give you an input, which is a bunch of text and some plus signs. I need you to extract all the drug names for me from each input, and output the corresponding list.\newline\newline
        Here are some of the input and output sample:\newline
        Input1: None.  +   Nifedipine XL, Calcitriol, Lisinopril, Aspirin, Lasix, Glyburide, Clonidine, Zoloft, Simvastatin, Tums, Procrit, Lupron, Niferex.\newline
        Output1: [Nifedipine XL, Calcitriol, Lisinopril, Aspirin, Lasix, Glyburide, Clonidine, Zoloft, Simvastatin, Tums, Procrit, Lupron, Niferex]\newline
        Input2: The patient was taking Aspirin, Atovaquone, Levofloxacin.  +  The patient was on multiple medications including Emtriva, Lisinoprol, Metoprolol, Stavudine.\newline
        Output2: [Aspirin, Atovaquone, Levofloxacin, Emtriva, Lisinoprol, Metoprolol, Stavudine]\newline\newline
        Now you need to provide the corresponding output of input3, without any other words:\newline
        Input3:  \{\{part 1\}\} + \{\{part 2\}\} + ... + \{\{part n\}\} \newline
        You only need to output a list!\newline
        Output3: \\
    \hline
    \end{tabular}
\end{table}


\begin{table}
    \centering
    \caption{Prompt template for evaluating GPT-3.5 (gpt-3.5-turbo-1106) and GPT-4 (gpt-4-0613) on medication recommendation, where the contents in ``\{\{\}\}'' are replaced by specific elements when fed to models.}
    \label{SI_tab:template_gpt}
    \begin{tabular}{p{0.95\textwidth}}
    \hline
        \rowcolor{mygray}
         \textbf{With history:} \\
         You are taking a medical exam right now. \\
         You will be given a patient's clinical condition with a candidate drug, your task is to judge whether the candidate drug is effective and safe for the patient. \\
         Answer with \textless Yes.\textgreater  or \textless No.\textgreater  and do not provide any other words. \\
         Do not repeat the question in your answer. \\
         \#\#\# Previous Hospitalization:  \{\{history\}\}  \\
         \#\#\# Current Clinical Condition:  \{\{input\}\}  \\
         \#\#\# Answer:  \{\{Yes. or No.\}\}\\
         
         \hline
         \rowcolor{mygray}
         \textbf{Without history:} \\
         You are taking a medical exam right now. \\
         You will be given a patient's clinical condition with a candidate drug, your task is to judge whether the candidate drug is effective and safe for the patient. \\
         Answer with \textless Yes.\textgreater  or \textless No.\textgreater  and do not provide any other words. \\
         Do not repeat the question in your answer. \\
         \#\#\# Current Clinical Condition: \{\{input\}\} \\
         \#\#\# Answer:  \{\{Yes. or No.\}\}\\
    \hline
    \end{tabular}
\end{table}

         
         

\begin{table}
    \centering
    \caption{Prompt template for evaluating MedAlpaca (medalpaca-13B) on medication recommendation, where the contents in ``\{\{\}\}'' are replaced by specific elements when fed to models.}
    \label{SI_tab:template_medalpaca}
    \begin{tabular}{p{0.95\textwidth}}
    \hline
        \rowcolor{mygray}
         \textbf{With history:} \\
         You will be given a patient's clinical condition with a candidate drug, your task is to judge whether the candidate drug is effective and safe for the patient. \\
         Do not repeat the question in your answer.\\
         \#\#\# Previous Hospitalization: \{\{history\}\}\\
         \#\#\# Current Clinical Condition: \{\{input\}\}\\
         \#\#\# Choose the appropriate option from the following two options: \textless Yes.\textgreater  or \textless No.\textgreater \\
         \#\#\# Answer:  \{\{Yes. or No.\}\}\\
         
         \hline
         \rowcolor{mygray}
         \textbf{Without history:} \\
         You will be given a patient's clinical condition with a candidate drug, your task is to judge whether the candidate drug is effective and safe for the patient. \\
         Do not repeat the question in your answer.\\
         \#\#\# Current Clinical Condition: \{\{input\}\}  \\
         \#\#\# Choose the appropriate option from the following two options: \textless Yes.\textgreater  or \textless No.\textgreater \\
         \#\#\# Answer: \\
    \hline
    \end{tabular}
\end{table}

\begin{table}
    \centering
    \caption{Prompt template for evaluating PMC-LLaMA (PMC\_LLaMA\_13B) on medication recommendation, where the contents in ``\{\{\}\}'' are replaced by specific elements when fed to models.}
    \label{SI_tab:template_PMC}
    \begin{tabular}{p{0.95\textwidth}}
    \hline
        \rowcolor{mygray}
         \textbf{With history:} \\
         You are taking a medical exam right now. \\
         You will be given a patient's clinical condition with a candidate drug, \\
         You need to determine whether the drug is appropriate, appropriate, and necessary for the patient. \\
         Only necessary, safe, and appropriate drugs should answer \textless Yes.\textgreater . \\
         Answer with \textless Yes.\textgreater  or \textless No.\textgreater  and do not provide any other words, do not explain anything.\\
         \#\#\# Previous Hospitalization: 
         \{\{history\}\}\\
         \#\#\# Current Clinical Condition:
         \{\{input\}\}\\
         \#\#\# Answer:  \{\{Yes. or No.\}\}\\
         
         \hline
         \rowcolor{mygray}
         \textbf{Without history:} \\
         You are taking a medical exam right now. \\
         You will be given a patient's clinical condition with a candidate drug, \\
         You need to determine whether the drug is appropriate, appropriate, and necessary for the patient. \\
         Only necessary, safe, and appropriate drugs should answer \textless Yes.\textgreater . \\
         Answer with \textless Yes.\textgreater  or \textless No.\textgreater  and do not provide any other words, do not explain anything.\\
         \#\#\# Current Clinical Condition:
         \{\{input\}\}\\
         \#\#\# Answer:  \{\{Yes. or No.\}\}\\
         
    \hline
    \end{tabular}
\end{table}

\begin{table}
    \centering
    \caption{Prompt template for evaluating Qwen (Qwen\_14B\_Chat) on medication recommendation, where the contents in ``\{\{\}\}'' are replaced by specific elements when fed to models.}
    \label{SI_tab:template_Qwen}
    \begin{tabular}{p{0.95\textwidth}}
    \hline
        \rowcolor{mygray}
         \textbf{With history:} \\
         You are taking a medical exam right now. You will be given a patient's clinical condition with a candidate drug, your task is to judge whether the candidate drug is effective and safe for the patient. Answer with \textless Yes.\textgreater  or \textless No.\textgreater  and do not provide any other words. Do not repeat the question in your answer. \\
         \#\#\# Previous Hospitalization:
         \{\{history\}\}\\
         \#\#\# Current Clinical Condition: 
         \{\{input\}\}\\
         \#\#\# Answer:  \{\{Yes. or No.\}\}\\
         
         \hline
         \rowcolor{mygray}
         \textbf{Without history:} \\
         You are taking a medical exam right now. You will be given a patient's clinical condition with a candidate drug, your task is to judge whether the candidate drug is effective and safe for the patient. Answer with \textless Yes.\textgreater  or \textless No.\textgreater  and do not provide any other words. Do not repeat the question in your answer.\\
         \#\#\# Current Clinical Condition: 
         \{\{input\}\}\\
         \#\#\# Answer:  \{\{Yes. or No.\}\}\\
         
    \hline
    \end{tabular}
\end{table}

\begin{table}
    \centering
    \caption{Prompt template for evaluating ChatGLM3 (ChatGLM3-6B) on medication recommendation, where the contents in ``\{\{\}\}'' are replaced by specific elements when fed to models.}
    \label{SI_tab:template_chatglm}
    \begin{tabular}{p{0.95\textwidth}}
    \hline
        \rowcolor{mygray}
         \textbf{With history:} \\
         You are taking a medical exam right now. \\
         You will be given a patient's clinical condition with a candidate drug delimited by triple quotes, \\
         your task is to judge whether the candidate drug is effective and safe for the patient. \\
         Note: Answer with \textless Yes.\textgreater  or \textless No.\textgreater  and do not explain.\\
         \#\#\# Previous Hospitalization: 
         \{\{history\}\}\\
         \#\#\# Current Clinical Condition:
         \{\{input\}\}\\
         \#\#\# Answer:  \{\{Yes. or No.\}\}\\
         
         \hline
         \rowcolor{mygray}
         \textbf{Without history:} \\
         You are taking a medical exam right now. \\
         You will be given a patient's clinical condition with a candidate drug delimited by triple quotes, \\
         your task is to judge whether the candidate drug is effective and safe for the patient. \\
         Note: Answer with \textless Yes.\textgreater  or \textless No.\textgreater  and do not explain.\\
         \#\#\# Current Clinical Condition: 
         \{\{input\}\}\\
         \#\#\# Answer:  \{\{Yes. or No.\}\}\\
         
    \hline
    \end{tabular}
\end{table}

\begin{table}
    \centering
    \caption{Prompt template for evaluating LLaMA-2 (Llama-2-7B) and LLaMA-3 (Llama-3-8B) on medication recommendation, where the contents in ``\{\{\}\}'' are replaced by specific elements when fed to models.}
    \label{SI_tab:template_llama}
    \begin{tabular}{p{0.95\textwidth}}
    \hline
        \rowcolor{mygray}
         \textbf{With history:} \\
         You are about to evaluate a candidate drug for a patient's clinical condition. You will be provided with the patient's current condition, as well as information about their previous hospitalization, and the candidate drug. Your task is to determine whether the candidate drug is effective and safe for the patient. Please respond with \textless Yes.\textgreater  or \textless No.\textgreater  without providing an explanation. \\
         \#\#\# Previous Hospitalization: 
         \{\{history\}\}\\
         \#\#\# Current Clinical Condition:
         \{\{input\}\}\\
         \#\#\# Answer:  \{\{Yes. or No.\}\}\\
         
         \hline
         \rowcolor{mygray}
         \textbf{Without history:} \\
         You are about to evaluate a candidate drug for a patient's clinical condition. You will be provided with the patient's current condition, as well as information about their previous hospitalization, and the candidate drug. Your task is to determine whether the candidate drug is effective and safe for the patient. Please respond with \textless Yes.\textgreater  or \textless No.\textgreater  without providing an explanation. \\
         \#\#\# Current Clinical Condition: 
         \{\{input\}\}\\
         \#\#\# Answer:  \{\{Yes. or No.\}\}\\
         
    \hline
    \end{tabular}
\end{table}


\clearpage  


\begin{table}
    \centering
    \caption{Prompt template for evaluating GPT-3.5 (gpt-3.5-turbo-1106), GPT-4 (gpt-4-0613), LLaMA-2 (Llama-2-7B), and LLaMA-3 (Llama-3-8B) on medical knowledge answering. ``\{\{drug\}\}'' and ``\{\{disease\}\}'' are replaced by specific elements when fed to models.}
    \label{SI_tab:knowledge-gpt&llama}
    \begin{tabular}{p{0.95\textwidth}}
    \hline
         You are taking a medical exam right now.\\
         Your task is to ignore the individual physical condition of the patient and only consider the matching relationship between the drug and the disease.\\
         From the following options enclosed in single quotes, select the one that most appropriately describes the drug-disease relationship, without any other word.\\
         Choices: ``\{indication\}, \{contraindication\}, \{off-label use\}, \{unknown\}''\\
         Drug: \{\{drug\}\}.\\
         Disease: \{\{disease\}\}.\\
         You just need to choose an answer from the above choices to give without any explanation!\\
         The most appropriate option without explanation is:\\
    \hline
    \end{tabular}
\end{table}

\begin{table}
    \centering
    \caption{Prompt template for evaluating ChatGLM3 (ChatGLM3-6B) on medical knowledge answering. 
    ``\{\{drug\}\}'' and ``\{\{disease\}\}'' are replaced by specific elements when fed to models.}
    \label{SI_tab:knowledge-chatglm}
    \begin{tabular}{p{0.95\textwidth}}
    \hline
        From the following two options, find the best option that describes the relationship between drug \{\{drug\}\} and disease \{\{disease\}\}\\
        Choices: ``A.indication  B.contraindication  C.off-label use  D.unknown''\\
        Candidates: ``\textless A.\textgreater  , \textless B.\textgreater, \textless C.\textgreater or \textless D.\textgreater''\\
        Answer: output one answer from candidates without any other word, the best option is: \\
    \hline
    \end{tabular}
\end{table}

\begin{table}
    \centering
    \caption{Prompt template for evaluating PMC-LLaMA (PMC\_LLaMA\_13B) and MedAlpaca (medalpaca-13B) on medical knowledge answering. ``\{\{drug\}\}'' and ``\{\{disease\}\}'' are replaced by specific elements when fed to models.}
    \label{SI_tab:knowledge-PMC&medalpaca}
    \begin{tabular}{p{0.95\textwidth}}
    \hline
        From the following two options, find the best option that describes the relationship between drug \{\{drug\}\} and disease \{\{disease\}\}\\
        Choice:  ``A.indication,  B.contraindication,  C.off-label use,  D.unknown''\\
        Answer: \\
    \hline
    \end{tabular}
\end{table}

\begin{table}
    \centering
    \caption{Prompt template for evaluating Qwen (Qwen-14B-Chat) on medical knowledge answering. ``\{\{drug\}\}'' and ``\{\{disease\}\}'' are replaced by specific elements when fed to models.}
    \label{SI_tab:knowledge-qwen}
    \begin{tabular}{p{0.95\textwidth}}
    \hline
        You are taking a medical exam right now. There is a multiple-choice question for you to answer. Please note that you need to give an answer whether you know the question or not.\\
        Please select the most descriptive of the following options that describes the relationship between drug \{\{drug\}\} and disease \{\{disease\}\}\\
        You only need to consider the general adaptive relationship between the disease and the drug, not the specific individual patient condition.\\
        Please make sure to choose the most appropriate answer from choices, do not give any other explanation, and do not refuse to answer for any reason.\\
        Choice:  ``\{indication\}, \{contraindication\}, \{off-label use\}, \{unknown\}''\\
        Answer:  \\
    \hline
    \end{tabular}
\end{table}

\begin{table}
\centering
    \caption{Medical knowledge test result of GPT-4 (gpt-4-0613). 
    The four settings define the output space of the LLM: \\
    Setting 1: \{``Indication'', ``Contraindication''\}; \\
    Setting 2: \{``Indication'', ``Contraindication'', ``Unknown''\}; \\
    Setting 3: \{``Indication'', ``Contraindication'', ``Off-label use''\}; \\
    Setting 4: \{``Indication'', ``Contraindication'', ``Off-label use'', ``Unknown''\}.
    }
    \label{SI_tab:knowledge_result-gpt4}
    \begin{tabular}{llllll}
    \hline
    \multirow{2}{*}{Setting}   & \multicolumn{1}{c}{\multirow{2}{*}{Ground-truth}} & \multicolumn{4}{c}{Output}                                 \\  \cmidrule{3-6} 
                               & \multicolumn{1}{c}{}                              & Indication    & Contraindication & Off-label use & Unknown \\ \hline
    \multirow{2}{*}{Setting 1} & Indication                                        & \textbf{88\%} & 12\%             & /             & /       \\
                               & Contraindication                                  & 6\%           & \textbf{94\%}    & /             & /       \\ \hline
    \multirow{2}{*}{Setting 2} & Indication                                        & \textbf{80\%} & 4\%              & /             & 16\%    \\
                               & Contraindication                                  & 3\%           & \textbf{38\%}    & /             & 59\%    \\ \hline
    \multirow{3}{*}{Setting 3} & Indication                                        & \textbf{78\%} & 7\%              & 15\%          & /       \\
                               & Contraindication                                  & 4\%           & \textbf{65\%}    & 31\%          & /       \\
                               & Off-label use                                     & 72\%          & 8\%              & \textbf{20\%} & /       \\ \hline
    \multirow{3}{*}{Setting 4} & Indication                                        & \textbf{74\%} & 4\%              & 2\%           & 20\%    \\
                               & Contraindication                                  & 5\%           & \textbf{34\%}    & 1\%           & 60\%    \\
                               & Off-label use                                     & 71\%          & 3\%              & \textbf{2\%}  & 24\%    \\ \hline
    \end{tabular}
\end{table}

\begin{table}
\centering
    \caption{Medical knowledge test result of GPT-3.5 (gpt-3.5-turbo-1106).
    The four settings define the output space of the LLM: \\
    Setting 1: \{``Indication'', ``Contraindication''\}; \\
    Setting 2: \{``Indication'', ``Contraindication'', ``Unknown''\}; \\
    Setting 3: \{``Indication'', ``Contraindication'', ``Off-label use''\}; \\
    Setting 4: \{``Indication'', ``Contraindication'', ``Off-label use'', ``Unknown''\}.
    }
    \label{SI_tab:knowledge_result-gpt3.5}
    \begin{tabular}{llllll}
        \hline
        \multirow{2}{*}{Setting}   & \multicolumn{1}{c}{\multirow{2}{*}{Ground-truth}} & \multicolumn{4}{c}{Output}                                 \\  \cmidrule{3-6} 
                                   & \multicolumn{1}{c}{}                              & Indication    & Contraindication & Off-label use & Unknown \\ \hline
        \multirow{2}{*}{Setting 1} & Indication & \textbf{97\%} & 3\% & / & / \\
                                   & Contraindication & 40\% & \textbf{60\%} & / & / \\ \hline

        \multirow{2}{*}{Setting 2} & Indication & \textbf{95\%} & 2\% & / & 3\% \\
                                   & Contraindication & 17\% & \textbf{43\%} & / & 40\% \\ \hline

        \multirow{3}{*}{Setting 3} & Indication & \textbf{96\%} & 1\% & 3\% & / \\
                                   & Contraindication & 33\% & \textbf{56\%} & 11\% & / \\
                                   & off-label use & 92\% & 6\% & \textbf{2\%} & / \\ \hline
        
        \multirow{3}{*}{Setting 4} & Indication & \textbf{92\%} & 3\% & 0\% & 5\% \\
                                   & Contraindication & 17\% & \textbf{36\%} & 0\% & 47\% \\
                                   & off-label use & 84\% & 4\% & \textbf{0\%} & 12\% \\ \hline
    \end{tabular}
\end{table}

\begin{table}
\centering
    \caption{Medical knowledge test result of LLaMA-2 (Llama-2-7B).
    The four settings define the output space of the LLM: \\
    Setting 1: \{``Indication'', ``Contraindication''\}; \\
    Setting 2: \{``Indication'', ``Contraindication'', ``Unknown''\}; \\
    Setting 3: \{``Indication'', ``Contraindication'', ``Off-label use''\}; \\
    Setting 4: \{``Indication'', ``Contraindication'', ``Off-label use'', ``Unknown''\}.}
    \label{SI_tab:knowledge_result-llama}
    \begin{tabular}{llllll}
    \hline
    \multirow{2}{*}{Setting}   & \multicolumn{1}{c}{\multirow{2}{*}{Ground-truth}} & \multicolumn{4}{c}{Output}                                 \\  \cmidrule{3-6} 
                               & \multicolumn{1}{c}{}                              & Indication    & Contraindication & Off-label use & Unknown \\ \hline
        \multirow{2}{*}{Setting 1} & Indication & \textbf{27\%} & 73\% & / & / \\
                                   & Contraindication & 18\% & \textbf{82\%} & / & / \\ \hline
        
        \multirow{2}{*}{Setting 2} & Indication & \textbf{25\%} & 14\% & / & 61\% \\
                                   & Contraindication & 8\% & \textbf{11\%} & / & 81\% \\ \hline
        
        \multirow{3}{*}{Setting 3} & Indication & \textbf{0\%} & 0\% & 100\% & / \\
                                   & Contraindication & 1\% & \textbf{1\%} & 98\% & / \\
                                   & off-label use & 4\% & 5\% & \textbf{91\%} & / \\ \hline

        \multirow{3}{*}{Setting 4} & Indication & \textbf{15\%} & 2\% & 68\% & 15\% \\
                                   & Contraindication & 7\% & \textbf{2\%} & 62\% & 29\% \\
                                   & off-label use & 11\% & 4\% & \textbf{72\%} & 13\% \\ \hline
        
    \end{tabular}
\end{table}

\begin{table}
\centering
    \caption{Medical knowledge test result of LLaMA-3 (Llama-3-8B).
    The four settings define the output space of the LLM: \\
    Setting 1: \{``Indication'', ``Contraindication''\}; \\
    Setting 2: \{``Indication'', ``Contraindication'', ``Unknown''\}; \\
    Setting 3: \{``Indication'', ``Contraindication'', ``Off-label use''\}; \\
    Setting 4: \{``Indication'', ``Contraindication'', ``Off-label use'', ``Unknown''\}.}
    \label{SI_tab:knowledge_result-llama3}
    \begin{tabular}{llllll}
    \hline
    \multirow{2}{*}{Setting}   & \multicolumn{1}{c}{\multirow{2}{*}{Ground-truth}} & \multicolumn{4}{c}{Output}                                 \\  \cmidrule{3-6} 
                               & \multicolumn{1}{c}{}                              & Indication    & Contraindication & Off-label use & Unknown \\ \hline
        \multirow{2}{*}{Setting 1} & Indication & \textbf{100\%} & 0\% & / & / \\
                                   & Contraindication & 88\% & \textbf{12\%} & / & / \\ \hline
        
        \multirow{2}{*}{Setting 2} & Indication & \textbf{100\%} & 0\% & / & 0\% \\
                                   & Contraindication & 91\% & \textbf{9\%} & / & 0\% \\ \hline
        
        \multirow{3}{*}{Setting 3} & Indication & \textbf{100\%} & 0\% & 0\% & / \\
                                   & Contraindication & 91\% & \textbf{9\%} & 0\% & / \\
                                   & off-label use & 100\% & 0\% & \textbf{0\%} & / \\ \hline

        \multirow{3}{*}{Setting 4} & Indication & \textbf{100\%} & 0\% & 0\% & 0\% \\
                                   & Contraindication & 88\% & \textbf{10\%} & 2\% & 0\% \\
                                   & off-label use & 100\% & 0\% & \textbf{0\%} & 0\% \\ \hline
        
    \end{tabular}
\end{table}

\begin{table}
\centering
    \caption{Medical knowledge test result of PMC-LLaMA (PMC\_LLaMA\_13B).
    The four settings define the output space of the LLM: \\
    Setting 1: \{``Indication'', ``Contraindication''\}; \\
    Setting 2: \{``Indication'', ``Contraindication'', ``Unknown''\}; \\
    Setting 3: \{``Indication'', ``Contraindication'', ``Off-label use''\}; \\
    Setting 4: \{``Indication'', ``Contraindication'', ``Off-label use'', ``Unknown''\}.}
    \label{SI_tab:knowledge_result-pmc}
    \begin{tabular}{llllll}
    \hline
    \multirow{2}{*}{Setting}   & \multicolumn{1}{c}{\multirow{2}{*}{Ground-truth}} & \multicolumn{4}{c}{Output}                                 \\  \cmidrule{3-6} 
                               & \multicolumn{1}{c}{}                              & Indication    & Contraindication & Off-label use & Unknown \\ \hline
        \multirow{2}{*}{Setting 1} & Indication & \textbf{79\%} & 21\% & / & / \\
                                   & Contraindication & 36\% & \textbf{64\%} & / & / \\ \hline
        
        \multirow{2}{*}{Setting 2} & Indication & \textbf{35\%} & 5\% & / & 60\% \\
                                   & Contraindication & 3\% & \textbf{9\%} & / & 88\% \\ \hline
        
        \multirow{3}{*}{Setting 3} & Indication & \textbf{83\%} & 3\% & 14\% & / \\
                                   & Contraindication & 37\% & \textbf{24\%} & 39\% & / \\
                                   & off-label use & 69\% & 8\% & \textbf{23\%} & / \\ \hline

        \multirow{3}{*}{Setting 4} & Indication & \textbf{73\%} & 7\% & 20\% & 0\% \\
                                   & Contraindication & 25\% & \textbf{26\%} & 49\% & 0\% \\
                                   & off-label use & 63\% & 7\% & \textbf{30\%} & 0\% \\ \hline
        
    \end{tabular}
\end{table}

\begin{table}
\centering
    \caption{Medical knowledge test result of MedAlpaca (medalpaca-13B).
    The four settings define the output space of the LLM: \\
    Setting 1: \{``Indication'', ``Contraindication''\}; \\
    Setting 2: \{``Indication'', ``Contraindication'', ``Unknown''\}; \\
    Setting 3: \{``Indication'', ``Contraindication'', ``Off-label use''\}; \\
    Setting 4: \{``Indication'', ``Contraindication'', ``Off-label use'', ``Unknown''\}.
    }
    \label{SI_tab:knowledge_result-medalpaca}
    \begin{tabular}{llllll}
    \hline
    \multirow{2}{*}{Setting}   & \multicolumn{1}{c}{\multirow{2}{*}{Ground-truth}} & \multicolumn{4}{c}{Output}                                 \\  \cmidrule{3-6} 
                               & \multicolumn{1}{c}{}                              & Indication    & Contraindication & Off-label use & Unknown \\ \hline
        \multirow{2}{*}{Setting 1} & Indication & \textbf{32\%} & 68\% & / & / \\
                                   & Contraindication & 4\% & \textbf{96\%} & / & / \\ \hline
        
        \multirow{2}{*}{Setting 2} & Indication & \textbf{61\%} & 31\% & / & 8\% \\
                                   & Contraindication & 18\% & \textbf{77\%} & / & 5\% \\ \hline
        
        \multirow{3}{*}{Setting 3} & Indication & \textbf{40\%} & 58\% & 2\% & / \\
                                   & Contraindication & 9\% & \textbf{87\%} & 4\% & / \\
                                   & off-label use & 25\% & 75\% & \textbf{0\%} & / \\ \hline

        \multirow{3}{*}{Setting 4} & Indication & \textbf{40\%} & 58\% & 2\% & 0\% \\
                                   & Contraindication & 9\% & \textbf{87\%} & 4\% & 0\% \\
                                   & off-label use & 25\% & 75\% & \textbf{0\%} & 0\% \\ \hline
        
    \end{tabular}
\end{table}

\begin{table}
\centering
    \caption{Medical knowledge test result of Qwen (Qwen-14B-Chat).
    The four settings define the output space of the LLM: \\
    Setting 1: \{``Indication'', ``Contraindication''\}; \\
    Setting 2: \{``Indication'', ``Contraindication'', ``Unknown''\}; \\
    Setting 3: \{``Indication'', ``Contraindication'', ``Off-label use''\}; \\
    Setting 4: \{``Indication'', ``Contraindication'', ``Off-label use'', ``Unknown''\}.
    }
    \label{SI_tab:knowledge_result-qwen}
    \begin{tabular}{llllll}
        \hline
        \multirow{2}{*}{Setting}   & \multicolumn{1}{c}{\multirow{2}{*}{Ground-truth}} & \multicolumn{4}{c}{Output}                                 \\  \cmidrule{3-6} 
                                  & \multicolumn{1}{c}{}                              & Indication    & Contraindication & Off-label use & Unknown \\ \hline
        \multirow{2}{*}{Setting 1} & Indication & \textbf{100\%} & 0\% & / & / \\
                                   & Contraindication & 74\% & \textbf{26\%} & / & / \\ \hline
        
        \multirow{2}{*}{Setting 2} & Indication & \textbf{90\%} & 0\% & / & 10\% \\
                                   & Contraindication & 32\% & \textbf{19\%} & / & 49\% \\ \hline
        
        \multirow{3}{*}{Setting 3} & Indication & \textbf{99\%} & 0\% & 1\% & / \\
                                   & Contraindication & 60\% & \textbf{35\%} & 5\% & / \\      
                                   & off-label use & 98\% & 2\% & \textbf{0\%} & / \\ \hline   

        \multirow{3}{*}{Setting 4} & Indication & \textbf{87\%} & 0\% & 0\% & 13\% \\
                                   & Contraindication & 25\% & \textbf{14\%} & 0\% & 61\% \\
                                   & off-label use & 81\% & 1\% & \textbf{0\%} & 18\% \\ \hline
        
    \end{tabular}
\end{table}

\begin{table}
\centering
    \caption{Medical knowledge test result of ChatGLM3 (ChatGLM3-6B).
    The four settings define the output space of the LLM: \\
    Setting 1: \{``Indication'', ``Contraindication''\}; \\
    Setting 2: \{``Indication'', ``Contraindication'', ``Unknown''\}; \\
    Setting 3: \{``Indication'', ``Contraindication'', ``Off-label use''\}; \\
    Setting 4: \{``Indication'', ``Contraindication'', ``Off-label use'', ``Unknown''\}.
    }
    \label{SI_tab:knowledge_result-chatglm}
    \begin{tabular}{llllll}
    \hline
    \multirow{2}{*}{Setting}   & \multicolumn{1}{c}{\multirow{2}{*}{Ground-truth}} & \multicolumn{4}{c}{Output}                                 \\  \cmidrule{3-6} 
                               & \multicolumn{1}{c}{}                              & Indication    & Contraindication & Off-label use & Unknown \\ \hline
        \multirow{2}{*}{Setting 1} & Indication & \textbf{88\%} & 12\% & / & / \\
                                   & Contraindication & 82\% & \textbf{18\%} & / & / \\ \hline
        
        \multirow{2}{*}{Setting 2} & Indication & \textbf{76\%} & 24\% & / & 0\% \\
                                   & Contraindication & 58\% & \textbf{39\%} & / & 3\% \\ \hline
        
        \multirow{3}{*}{Setting 3} & Indication & \textbf{82\%} & 12\% & 6\% & / \\
                                   & Contraindication & 68\% & \textbf{30\%} & 2\% & / \\
                                   & off-label use & 71\% & 22\% & \textbf{7\%} & / \\ \hline

        \multirow{3}{*}{Setting 4} & Indication & \textbf{66\%} & 2\% & 11\% & 21\% \\
                                   & Contraindication & 65\% & \textbf{4\%} & 1\% & 30\% \\
                                   & off-label use & 60\% & 4\% & \textbf{10\%} & 26\% \\ \hline
        
    \end{tabular}
\end{table}



\begin{table}
\centering
\caption{Brief descriptions of compared methods}
\label{tab:method_descriptions}
\begin{tabularx}{1.15\textwidth}{lX}  
\toprule
\textbf{Method} & \textbf{Description} \\
\midrule
\textbf{LR} & A standard logistic regression that formulates medication recommendation as multi-label classification using multi-hot inputs. \\
\textbf{LEAP}~[1] & An instance-based method using LSTM to generate medication sequences. \\
\textbf{GAMENet}~[2] & Integrates memory networks and graph convolution to model historical EHR data. \\
\textbf{SafeDrug}~[3] & Uses drug molecular graphs to ensure safe medication recommendations. \\
\textbf{COGNet}~[4] & Applies a copy-or-predict mechanism in an encoder-decoder framework for sequential drug prediction. \\
\textbf{MICRON}~[5] & A recurrent residual learning model considering medication changes and past combinations. \\
\textbf{MoleRec}~[6] & Models interactions between molecular substructures and health conditions. \\
\textbf{LEADER}~[7] & Transforms inputs into text and distills knowledge from LLMs for semantic understanding. \\
\textbf{RAREMed}~[8] & Improves code representation for rare diseases using a pretrain-finetune framework. \\
\textbf{ChatGLM3}~[9] & A bilingual GLM-based LLM with 6.2B parameters, trained on Chinese-English data and optimized for QA and dialogue. \\
\textbf{Qwen}~[10] & An open-source model by Alibaba, trained on web text, books, and code. \\
\textbf{LLaMA-2}~[11] & Meta's 7B-parameter Transformer LLM trained on 2T tokens of public data. \\
\textbf{LLaMA-3}~[12] & An 8B-parameter model from Meta, fine-tuned with SFT and RLHF for improved alignment and dialogue performance.\\
\textbf{GPT-3.5/4}~[13,14] & Commercial OpenAI models with strong general and medical NLP capabilities. \\
\textbf{MedAlpaca}~[15] & A medical version of Alpaca fine-tuned on instruction-following dialogues and QA. \\
\textbf{PMC-LLaMA}~[16] & A medical LLM fine-tuned on LLaMA using biomedical papers, textbooks, and domain instructions. \\
\bottomrule
\end{tabularx}
\begin{enumerate}[label={[\arabic*]}] \footnotesize
\item Zhang, Y., Chen, R., Tang, J., Stewart, W.F., Sun, J.: Leap: learning to prescribe effective and safe treatment combinations for multimorbidity. In: Proceedings of the 23rd ACM SIGKDD International Conference on Knowledge Discovery and Data Mining, pp. 1315–1324 (2017)

\item Shang, J., Xiao, C., Ma, T., Li, H., Sun, J.: GAMENet: Graph augmented memory networks for recommending medication combination. In: Proceedings of the AAAI Conference on Artificial Intelligence, vol. 33, pp. 1126–1133 (2019)

\item Yang, C., Xiao, C., Ma, F., Glass, L., Sun, J.: SafeDrug: Dual molecular graph encoders for recommending effective and safe drug combinations. arXiv preprint arXiv:2105.02711 (2021)

\item Wu, R., Qiu, Z., Jiang, J., Qi, G., Wu, X.: Conditional Generation Net for Medication Recommendation. In: Proceedings of the ACM Web Conference 2022, pp. 935–945 (2022)

\item Yang, C., Xiao, C., Glass, L., Sun, J.: Change matters: Medication change prediction with recurrent residual networks. arXiv preprint arXiv:2105.01876 (2021)

\item Yang, N., Zeng, K., Wu, Q., Yan, J.: MoleRec: Combinatorial drug recommendation with substructure-aware molecular representation learning. In: Proceedings of the ACM Web Conference 2023, pp. 4075–4085 (2023)

\item Liu, Q., Wu, X., Zhao, X., Zhu, Y., Zhang, Z., Tian, F., Zheng, Y.: Large language model distilling medication recommendation model. arXiv preprint arXiv:2402.02803 (2024)

\item Zhao, Z., Jing, Y., Feng, F., Wu, J., Gao, C., He, X.: Leave no patient behind: Enhancing medication recommendation for rare disease patients. arXiv preprint arXiv:2403.17745 (2024)

\item Du, Z., Qian, Y., Liu, X., Ding, M., Qiu, J., Yang, Z., Tang, J.: GLM: General language model pretraining with autoregressive blank infilling. In: Proceedings of the 60th Annual Meeting of the Association for Computational Linguistics (Volume 1: Long Papers), pp. 320–335 (2022)

\item Bai, J., Bai, S., Chu, Y., Cui, Z., Dang, K., Deng, X., Fan, Y., Ge, W., Han, Y., Huang, F., et al.: Qwen Technical Report. arXiv preprint arXiv:2309.16609 (2023)

\item Touvron, H., Martin, L., Stone, K., Albert, P., Almahairi, A., Babaei, Y., Bashlykov, N., Batra, S., Bhargava, P., Bhosale, S., et al.: LLaMA 2: Open foundation and fine-tuned chat models. arXiv preprint arXiv:2307.09288 (2023)

\item Grattafiori, A., Dubey, A., Jauhri, A., Pandey, A., Kadian, A., Al-Dahle, A., Letman, A., Mathur, A., Schelten, A., Vaughan, A., et al.: The LLaMA 3 herd of models. arXiv preprint arXiv:2407.21783 (2024)

\item Brown, T., Mann, B., Ryder, N., Subbiah, M., Kaplan, J.D., Dhariwal, P., Neelakantan, A., Shyam, P., Sastry, G., Askell, A., et al.: Language models are few-shot learners. Advances in Neural Information Processing Systems 33, 1877–1901 (2020)

\item Achiam, J., Adler, S., Agarwal, S., Ahmad, L., Akkaya, I., Aleman, F.L., Almeida, D., Altenschmidt, J., Altman, S., Anadkat, S., et al.: GPT-4 Technical Report. arXiv preprint arXiv:2303.08774 (2023)

\item Han, T., Adams, L.C., Papaioannou, J.-M., Grundmann, P., Oberhauser, T., Löser, A., Truhn, D., Bressem, K.K.: MedAlpaca – An open-source collection of medical conversational AI models and training data. arXiv preprint arXiv:2304.08247 (2023)

\item Wu, C., Zhang, X., Zhang, Y., Wang, Y., Xie, W.: PMC-LLAMA: Further finetuning LLaMA on medical papers. arXiv preprint arXiv:2304.14454 (2023)
\end{enumerate}
\end{table}
\begin{table}
\centering
\caption{Medication recommendation accuracy across various models.}
\label{tab:overall}
\begin{tabular}{lcccccc}
\hline
Model name & F1 & Jaccard & Recall & Precision & \#Med 
\\ \hline

\rowcolor{mygray}
\multicolumn{6}{l}{General LLMs} \\ 
ChatGLM & 0.0015 & 0.0008 & 0.0008 & 0.0125 & 0.15 & \\
Qwen & 0.1825 & 0.1094 & 0.315 & 0.2337 & 35.13 &
\\
LLaMA-2 & 0.2802 & 0.1662 & 0.9890 & 0.1665 & 149.43
\\
LLaMA-3 & 0.2805 & 0.1663 & 0.9994 & 0.1664 & 150.94
\\
GPT-3.5 & 0.3402 & 0.2109 & 0.5916 & 0.2650 & 54.46 &  \\
GPT-4  & 0.3542 & 0.2200 & 0.8249 & 0.2347 & 85.83 &  \\ 
\hline

\rowcolor{mygray}
\multicolumn{6}{l}{Medical LLMs} \\ 

MedAlpaca & 0.2723 & 0.1613 & 0.9230 & 0.1746 & 139.01 
\\
PMC-LLaMA & 0.2351 & 0.1371 & 0.5837 & 0.2121 & 82.96 \\
 \hline

\rowcolor{mygray}
\multicolumn{6}{l}{Conventional Models} \\ 

LR & 0.4928 & 0.3397 & 0.4365 & 0.6192 & 16.68 &  \\
LEAP & 0.4588 & 0.3054 & 0.4594 & 0.5050 & 19.73 &  \\
GAMENet & 0.5398 & 0.3829 & 0.5193 & 0.6013 & 19.73 &  \\
SafeDrug & 0.5037 & 0.3461 & 0.5226 & 0.5161 & 22.70 &  \\
COGNet & 0.5401 & 0.3822 & 0.6392 & 0.4956 & 29.35 &  \\
MICRON & 0.5419 & 0.3843 & 0.5401 & 0.5863 & 20.49 &  \\    
MoleRec & 0.5684 & 0.4089 & 0.6158 & 0.5558 & 25.15 &  \\
LEADER & 0.5350 & 0.3786 & 0.4648 & 0.6798 & 16.02 &  \\ 
RAREMed & 0.5794 & 0.4194 & 0.6063 & 0.5798 & 23.49 &  \\ 
\hline
LAMO (Ours) & 0.6294 & 0.4701 & 0.6485 & 0.6372 & 23.66 &  \\ \hline
\end{tabular}
\footnotesize
    A higher F1 score, Jaccard index, Recall, and Precision indicate better accuracy in medication recommendations, while a smaller difference from the ground truth (\#Med) of 22.93 signifies better performance. The evaluation of performance metrics utilizes various models, including the OpenAI API gpt-3.5-turbo-1106 for GPT-3.5, the OpenAI API gpt-4-0613 for GPT-4, Qwen-14B-Chat for Qwen, ChatGLM3-6B for ChatGLM, medalpaca-13B for MedAlpaca, PMC\_LLaMA\_13B for PMC-LLaMA, Llama-2-7B for LLaMA-2, and Llama-3-8B for LLaMA-3.
\end{table}

\begin{figure}[t]
    \centering
    \includegraphics[width=0.9\textwidth]{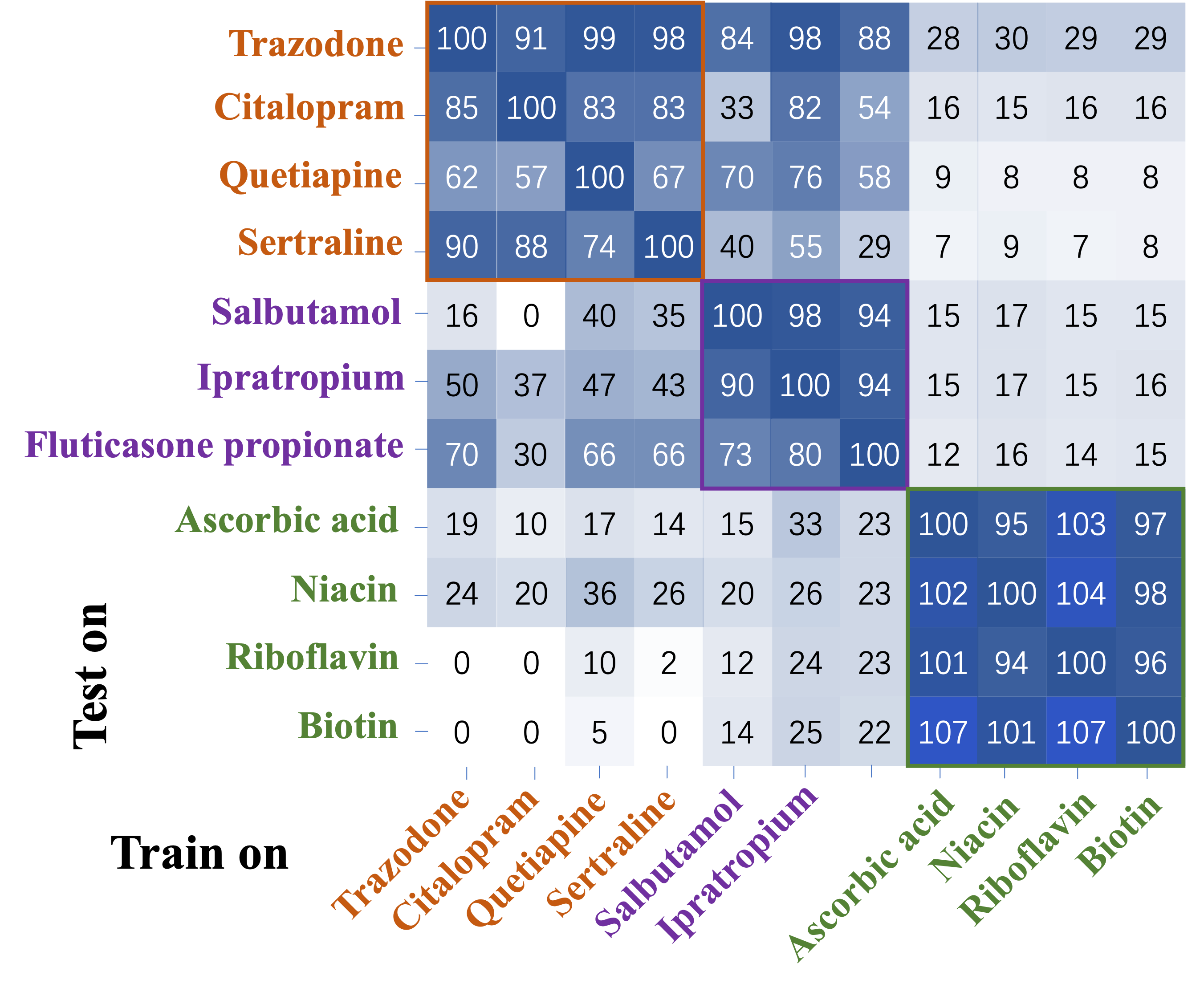}
    \caption{
    Out-of-distribution validation was conducted, where we present the relative performance (expressed as the percentage of zero-shot transfer performance compared to original performance) on target medications when trained on source medications. Medication names are color-coded to denote their types: orange for psychotropic drugs, purple for bronchodilators, and green for nutritional supplements.
    }
    \label{fig:generalization}
\end{figure}


\begin{table}[ht]
\centering
\begin{minipage}{0.98\textwidth}
{
\normalsize
\justifying
\noindent
\textbf{Table 25}: Summary statistics of the processed datasets and prompts. For MIMIC-III, we followed standard preprocessing and randomly split the data into training, validation, and test sets (4:1:1). Table~\ref{SI_tab:data_statistic} summarizes the structured data, and Table~\ref{SI_tab:statistic_notes_prompt} provides statistics of extracted note segments and prompts. For temporal validation, we applied the same processing pipeline to MIMIC-IV and divided it into four time intervals using the ``anchor year group''. Tables~\ref{SI_tab:data_statistic_2} and~\ref{SI_tab:statistic_notes_prompt_2} present the corresponding statistics for structured data and clinical notes.
}
\end{minipage}
\label{tab:combined_stats}

\vspace{0.5em}  

\begin{subtable}{0.8\textwidth}
    \centering
    \caption{Statistics of processed data in MIMIC-III}
    \label{SI_tab:data_statistic}
    \begin{tabular}{l|c}
        \hline
        Items & MIMIC-III \\
        \hline
        \# of visits / \# of patients & 14207 / 6226 \\
        dis. / prod. / med. space size & 1676 / 511 / 151 \\
        avg. / max \# of visits & 2.28 / 29 \\
        avg. / max \# of dis. per visit & 13.59 / 39 \\
        avg. / max \# of prod. per visit & 4.23 / 27 \\
        avg. / max \# of med. per visit & 23.36 / 77 \\
        \hline
    \end{tabular}
\end{subtable}


\begin{subtable}{0.95\textwidth}
    \centering
    \caption{Statistics of extracted clinical note fields and prompt in MIMIC-III}
    \label{SI_tab:statistic_notes_prompt}
    \begin{tabular}{l|c|l|c}
        \hline
         & avg. / max Value &  & avg. / max Value \\
        \hline
        \# of len(HPI) & 302.64 / 2527 & \# of HPI tokens   & 66.67 / 570 \\
        \# of len(PMH) & 275.87 / 1510 & \# of PMH tokens   & 65.05 / 356 \\
        \# of allergies & 1.90 / 28 & - & - \\
        \# of MoA & 10.02 / 52 & - & - \\
        \hline
        \# of len(note) & 813.52 / 3442 & \# of note tokens   & 205.95 / 842 \\
        \# of len(prompt) & 2205.65 / 5679 & \# of prompt tokens   & 591 / 1600 \\
        \hline
    \end{tabular}
\end{subtable}


\begin{subtable}{0.95\textwidth}
    \centering
    \caption{Statistics of processed MIMIC-IV structured data by year interval}
    \label{SI_tab:data_statistic_2}
    \begin{tabular}{l|cc}
    \hline
      Items & 2008-2010 & 2011-2013 \\
    \hline
    \# of visits / \# of patients     & 5915 / 2262 & 2772 / 1220 \\
    dis. / prod. / med. space size    & 1222 / 310 / 128 & 1198 / 300 / 128 \\
    avg. / max \# of visits           & 2.61 / 41 & 2.27 / 26 \\
    avg. / max \# of dis. per visit   & 11.74 / 38 & 11.72 / 37 \\
    avg. / max \# of prod. per visit  & 2.33 / 17 & 2.21 / 15 \\
    avg. / max \# of med. per visit   & 14.37 / 74 & 13.57 / 53 \\
    \hline
      Items & 2014-2016 & 2017-2019 \\
    \hline
    \# of visits / \# of patients     & 1917 / 964 & 1009 / 538 \\
    dis. / prod. / med. space size    & 1181 / 291 / 125 & 496 / 95 / 120 \\
    avg. / max \# of visits           & 1.99 / 21 & 1.88 / 13 \\
    avg. / max \# of dis. per visit   & 12.75 / 35 & 13.03 / 35 \\
    avg. / max \# of prod. per visit  & 2.17 / 20 & 1.99 / 11 \\
    avg. / max \# of med. per visit   & 11.93 / 52 & 9.47 / 39 \\
    \hline
    \end{tabular}
\end{subtable}

\begin{subtable}{0.95\textwidth}
    \centering
    \caption{Statistics of extracted clinical note fields and prompt in MIMIC-IV by year interval}
    \label{SI_tab:statistic_notes_prompt_2}
    \begin{tabular}{l|cccc}
    \hline
      Items & 2008-2010 & 2011-2013 & 2014-2016 & 2017-2019  \\
    \hline
     \# of len(HPI) & 181.07 / 942 & 179.28 / 899 & 185.21 / 1399 & 201.07 / 1056 \\
     \# of HPI tokens & 39.99 / 205 & 39.72 / 224 & 41.50 / 395 & 45.38 / 245 \\
     \# of len(PMH) & 210.49 / 617 & 185.97 / 670 & 192.17 / 639 & 198.36 / 706 \\
     \# of PMH tokens & 47.72 / 134 & 44.37 / 152 & 46.42 / 160 & 47.81 / 172 \\
     \# of allergies & 2.14 / 20 & 1.87 / 21 & 1.88 / 20 & 1.654 / 12 \\
     \# of MoA & 8.57 / 42 & 7.60 / 29 & 8.05 / 37 & 8.626 / 37 \\
      \hline
     \# of len(note) & 618.10 / 2011 & 586.11 / 1528 & 613.96 / 1740 & 652.26 / 1604 \\
     \# of note tokens & 162.82 / 504 & 152.53 / 394 & 161.03 / 483 & 171.06 / 476 \\
     \# of len(prompt) & 1479.24 / 4478 & 1402.06 / 3503 & 1436.25 / 3999 & 1396.28 / 3725 \\
     \# of prompt tokens & 384.25 / 1257 & 361.28 / 961 & 367.94 / 1103 & 352.98 / 933 \\
    
    \hline
    \end{tabular}
\end{subtable}
\end{table}

\end{document}